\documentclass[onecolumn]{svjour3}          

\smartqed  
\usepackage{graphicx}
\usepackage{amsmath}
\begin{document}

\title{Contribution of pressure to the energy-momentum density in a moving perfect fluid: A physical perspective} 
\titlerunning{Contribution of pressure to energy-momentum density}  
\author{Ashok K. Singal}
\authorrunning{A. K. Singal} 
\institute{A. K. Singal \at
Astronomy and Astrophysics Division, Physical Research Laboratory,
Navrangpura, Ahmedabad - 380 009, India\\
\email{ashokkumar.singal@gmail.com}}
\date{Received: date / Accepted: date}
\maketitle
\begin{abstract}
In the energy-momentum density expressions for a relativistic perfect fluid with a bulk motion, one comes across a couple of pressure-dependent terms, which though well known, are to an extent, lacking in their conceptual basis and the ensuing physical interpretation. In the expression for the energy density, the rest mass density along with the kinetic energy density of the fluid constituents due to their random motion, which contributes to the pressure as well, are already included. However, in a fluid with a bulk motion, there are, in addition, a couple of explicit, pressure-dependent terms in the energy-momentum densities, whose presence to an extent, is shrouded in mystery, especially from a physical perspective. We show here that one such pressure-dependent term appearing in the energy density, represents the work done by the fluid pressure against the Lorentz contraction during transition from the rest frame of the fluid to another frame in which the fluid has a bulk motion. This applies equally to the electromagnetic energy density of electrically charged systems in motion and explains in a natural manner an  apparently paradoxical result that the field energy of a charged capacitor system decreases with an increase in the system velocity. The momentum density includes another pressure-dependent term, that represents an energy flow across the system, due to the opposite signs of work being done by pressure on two opposite sides of the moving fluid. From Maxwell's stress tensor we demonstrate that in the expression for  electromagnetic momentum  of an electric charged particle, it is the presence of a similar pressure term,  arising from  electrical self-repulsion forces in the charged sphere, that yields a natural solution for the notorious, more than a century old but thought by many as still unresolved, 4/3 problem in the electromagnetic momentum.
\end{abstract}
\section{Introduction}
Recently it was shown that the energy-momentum densities of an ideal gas,  which is a simple example of a perfect fluid, can be derived, from first principles, in an inertial frame where the fluid possesses a bulk motion \cite{85}. One starts from the simple expressions for the energy density and pressure of an ideal gas in the rest frame of the gas, where the gas molecules may possess a random motion, but no bulk motion, and where the fluid is described by only the energy density and pressure. From a Lorentz transformation of the velocity vectors of molecules, moving along different directions in the rest frame of the fluid, their energy-momentum vectors and number densities are computed and from that one arrives at the energy-momentum density of the fluid in a frame where it has a bulk motion. 

In the energy-momentum tensor of a perfect fluid undergoing a bulk motion, one comes across a couple of pressure-dependent terms, making explicit contributions to energy and momentum densities. Mathematically speaking, the genesis of these terms is well understood, however, their conceptual basis and the ensuing physical interpretation remains shrouded in mystery, and it is not evident how the pressure within the medium gives rise to these terms.
Here we shall endeavour to explain conceptually how these terms come into play in the presence of pressure in the system. 

As we will show, the work done by the fluid pressure against the Lorentz contraction during transition from the rest frame of the fluid to a frame in which the fluid has a bulk motion, is one such factor. Another pressure-dependent term, contributing to the momentum density, arises from the opposite amounts of work being done by pressure on two opposite sides of the moving fluid, resulting in a flow of energy across the system, between these opposite sides. As we will show, the latter term could be relevant even when the bulk motion of the fluid could be non-relativistic in nature and it is one such term that has given rise to the famous 4/3 factor \cite{29,1,2} in the electromagnetic momentum of a moving charge, a problem more than a century old but on whose proper resolution the opinions still differ \cite{1,2,15,31,Rohrlich60}. Further, these terms are also responsible for the apparent paradox of a nil electromagnetic momentum in the case of a moving, electrically charged system, even when there might be a finite electromagnetic energy stored inside, like, e.g., in the case of a charged parallel-plate capacitor, moving in a direction perpendicular to the plate surfaces, with no magnetic field and hence no electromagnetic momentum in the system  \cite{15,31}. In order to eliminate these, apparently undesirable features of the electromagnetic theory, especially the factor of 4/3, a modified definition of the electromagnetic energy-momentum, has been put forward in the literature \cite {Rohrlich60}, including in many standard text-books \cite {1,2}. As we will demonstrate, there is no reason for adopting any such drastic changes in the conventional formulations. After all, no such modifications in the definitions of the energy-momentum of a system have ever been found to be necessary in the case of relativistic fluid models, and the physics remains the same in the electromagnetic case too.  

Unless otherwise specified we use cgs system of units throughout.
\section{Energy-momentum tensor of a perfect fluid} 
A perfect fluid is described completely in its comoving frame, called the rest frame of the fluid, by two quantities -- the energy density, a scalar, and the pressure, assumed to be isotropic \cite{MTW73,SC85,DS98}.
\begin{eqnarray}
\label{eq:84.1}
{\displaystyle {\cal T}^{\alpha' \beta' }=
 {\displaystyle {\begin{bmatrix}\rho_0&0&0&0\\0&p&0&0\\0&0&p&0\\0&0&0&p\end{bmatrix}}} =\operatorname {diag} (\rho_0 ,p,p,p)}\:,
\end{eqnarray}
where $\rho_0$ is the energy density and ${\displaystyle p}$ represents the pressure; both quantities are defined in the rest frame of the fluid. We follow here the convention where Greek letters $\alpha, \beta, \mu, \nu$ etc. take the values $0,1,2,3$ while Latin letters $i,j$ take the values $1,2,3$, a prime on an index indicating it refers to the rest frame of the fluid. Perfect fluids are supposed to have no heat conduction, no viscosity and no sheer stress, with ${\cal T}^{0'i'} = {\cal T}^{i'0'} = 0$ as well as ${\cal T}^{i'j'}= 0$ for $i \ne j$, i.e., all off-diagonal terms are zero in the rest frame.

The fluid has no bulk motion in the rest frame, though on micro scale there may be random motion of its constituents, which gives rise to pressure. The energy density $\rho_0 $ is the {\em total} energy per unit volume, e.g., in the ideal gas model of a fluid, it includes not only the rest mass energy of gas molecules, but also their kinetic energy that gives rise to the pressure. In case the fluid comprises electromagnetic radiation, as for instance in cosmological models \cite{MTW73,SC85}, then the energy density includes the radiation energy density as well. In fact, $\rho_0$ is sum of all possible contributions of the energy density to the fluid. Similar is the case for the pressure term, which includes contributions from {\em all} constituents, whether matter or radiation, that make the fluid.

Let ${\cal S}_0$ be the the rest frame of the fluid, which moves with velocity ${\bf v}$ along the x-axis with respect to the lab frame ${\cal S}$. The matrix for Lorentz transformation from,  frame ${\cal S}_0$ to the lab frame ${\cal S}$ is written as \cite{MTW73}
\begin{eqnarray}
\label{eq:84.1a}
{\displaystyle \Lambda^{\mu}_{\alpha'}=
 {\displaystyle {\begin{bmatrix}\gamma&\gamma v/c&0&0\\\gamma v/c&\gamma&0&0\\0&0&1&0\\0&0&0&1\end{bmatrix}}} }\:,
\end{eqnarray}
where $\gamma =1/\sqrt{1-v^{2}/c^{2}}$ is the Lorentz factor.

The energy-momentum tensor, accordingly, transforms as
\begin{eqnarray}
\label{eq:84.1b}
{\displaystyle {\cal T}^{\mu\nu}={\Lambda ^{\mu}}_{\alpha'}{\Lambda ^{\nu}}_{\beta'}{\cal T}^{\alpha' \beta'}.} 
\end{eqnarray}

From this, the energy-momentum tensor of a perfect fluid can be written in a frame-independent form \cite{MTW73,SC85,DS98}
\begin{eqnarray}
\label{eq:84.2}
{\displaystyle {\cal T}^{\mu \nu }=(\rho_0+{p})\,\frac{u^{\mu }u^{\nu }}{c^2}+p\,\eta ^{\mu \nu }\,}\:,
\end{eqnarray}
where u is the 4-velocity ($u^0=\gamma c$, $u^i=\gamma v^i$) of the fluid and  
\begin{eqnarray}
\label{eq:84.3}
{\displaystyle \eta^{\mu \nu }=\eta _{\mu \nu} 
 =\operatorname {diag} (-1,1,1,1)}\:,
\end{eqnarray}
is the metric tensor \cite{MTW73,SC85} of the special relativity.  In the rest frame, ($u^0=c$, $u^i=0$), and the expression for energy-momentum tensor in equation~(\ref{eq:84.2}) reduces to the diagonal form of equation~(\ref{eq:84.1}).

We can express the energy-momentum tensor in terms of the components, to get
\begin{eqnarray}
\label{eq:84.4}
{\cal T}^{00} = \left(\rho_0+{p}\right){u^0u^0\over c^2} - p \:,
\end{eqnarray}
\begin{eqnarray}
\label{eq:84.5}
{\cal T}^{0i} = {\cal T}^{i0} = \left(\rho_0+{p}\right){u^0u^i\over c^2} \:,
\end{eqnarray}
\begin{eqnarray}
\label{eq:84.6}
{\cal T}^{ij} = \left(\rho_0+{p}\right) {u^iu^j\over c^2} + p \delta^{ij}\: .
\end{eqnarray}
Here ${\cal T}^{00}$ is the energy density, ${\cal T}^{0i}$ is $1/c$ times the energy flux (energy crossing a unit area per unit time) normal to $x_i$, ${\cal T}^{i0}$ is $c$ times the momentum density along direction $x_i$ and ${\cal T}^{ij}$ is the stress ($i$ component of the momentum flux, i.e., the momentum crossing the surface $x_j$=constant per unit area per unit time). The energy-momentum tensor is symmetric with ${\cal T}^{0i}={\cal T}^{i0}$ as well as ${\cal T}^{ij}={\cal T}^{ji}$.

Now, in the rest frame ${\cal S}_0$, $u^0=c$ and $u^i=0$, and various components of energy-momentum tensor are
\begin{eqnarray}
\label{eq:84.4b}
{\cal T}^{0'0'} = \rho_0\:,
\end{eqnarray}
\begin{eqnarray}
\label{eq:84.5b}
{\cal T}^{0'i'} = {\cal T}^{i'0'} = 0\:.
\end{eqnarray}
\begin{eqnarray}
\label{eq:84.6b}
{\cal T}^{1'1'} ={\cal T}^{2'2'}={\cal T}^{3'3'} =  p\: ,
\end{eqnarray}
all other, non-diagonal, stress terms being zero. 

However, for the lab frame ${\cal S}$, with $u^0=\gamma c$, $u^1=\gamma v, u^2=0, u^3=0$, and we obtain various components of the energy-momentum tensor as
\begin{eqnarray}
\label{eq:84.4c}
{\cal T}^{00} = \gamma^2 \left(\rho_0 +p {v^2\over c^2}\right)\:,
\end{eqnarray}
\begin{eqnarray}
\label{eq:84.5c}
{\cal T}^{10} = \gamma^2\left(\rho_0 +p\right) {v\over c}\: ,
\end{eqnarray}
\begin{eqnarray}
\label{eq:84.5c1}
{\cal T}^{20} = {\cal T}^{30} = 0 \:,
\end{eqnarray}
\begin{eqnarray}
\label{eq:84.6c}
{\cal T}^{11} =  \gamma^2 \left(\rho_0 +p\right){v^2\over c^2}+p
\:,
\end{eqnarray}
\begin{eqnarray}
\label{eq:84.6c1}
{\cal T}^{22}={\cal T}^{33} =  p \:,
\end{eqnarray}
with all other, non-diagonal, stress terms zero.
\subsection{Radiation as perfect fluid}
If we assume the perfect fluid to be comprising only pure radiation, with the relative density and pressure contribution of matter, if any, negligible, then knowing that for radiation \cite{85,MTW73} 
\begin{eqnarray}
\label{eq:84.4.1}
p=\rho_0/3,
\end{eqnarray}
we can write the fluid components in the lab frame ${\cal S}$ as
\begin{eqnarray}
\label{eq:84.4a}
{\cal T}^{00} =  \gamma^2 \rho_0\left(1 + {v^2\over 3c^2}\right)\:,
\end{eqnarray}
\begin{eqnarray}
\label{eq:84.5a}
{\cal T}^{10}= \frac{4}{3}\,\gamma^2\rho_0  {v\over c} \:,
\end{eqnarray}
\begin{eqnarray}
\label{eq:84.5a1}
{\cal T}^{20} = {\cal T}^{30} = 0 \:,
\end{eqnarray}
\begin{eqnarray}
\label{eq:84.6a}
{\cal T}^{11} =   \frac{4}{3}\gamma^2\rho_0 {v^2\over c^2}+\frac{\rho_0}{3} \:,
\end{eqnarray}
\begin{eqnarray}
\label{eq:84.6a1}
{\cal T}^{22}={\cal T}^{33} =  p=\frac{\rho_0}{3}\:,
\end{eqnarray}
with all non-diagonal stress terms zero.
\section{A puzzle -- why does the pressure term make an appearance in the energy momentum of a moving system?} 
Let us examine various terms in the expression for energy-momentum tensor in the lab frame ${\cal S}$. For definiteness, we assume that the perfect fluid is confined within a volume $V_0$ in the rest frame ${\cal S}_0$. For simplicity, we assume for our purpose, the fluid to be uniform within the volume in the rest frame  ${\cal S}_0$, where the volume is in the shape of a rectangular box, as shown in Fig.~1. It could even be spherical or any arbitrary shape, as long as the condition of the uniformity of fluid, and the isotropy of the pressure inside the volume is maintained. 

The stress tensor relations ${\cal T}^{ij}$ in equations~(\ref{eq:84.6c}) and (\ref{eq:84.6c1}) are easily understood from the pressure term $p$, and the momentum density term ${\cal T}^{10}$ in  equation~(\ref{eq:84.5c}). The pressure $p$, defined as force per unit area on a surface element stationary in the rest frame ${\cal S}_0$, is actually an invariant quantity, which could be understood this way. The force component along $\bf v$, on a surface element stationary in frame ${\cal S}_0$, remains unchanged during a Lorentz transformation to frame ${\cal S}$, or $F_\parallel=  F_{0\parallel}$ \cite{KKR73}. Also, the area of the surface element, normal to the direction of motion, remains unchanged during this transformation. Therefore pressure, force per unit area, also remains unchanged during transformation from ${\cal S}_0$ to ${\cal S}$. 
For the force component normal to the direction of motion, $F_\perp=  F_{0\perp}/\gamma$  \cite{KKR73}. But then even the corresponding area element, due to the Lorentz contraction, is smaller in 
${\cal S}$ by the same factor $\gamma$, ensuring thereby the force per unit area to remain  unchanged. Thus pressure is an invariant quantity. This explains appearance of $p$, in equation~(\ref{eq:84.6c1}) as well as the last term in equation~(\ref{eq:84.6c}), in a form unchanged from that in equation~(\ref{eq:84.6b}) for the rest frame ${\cal S}_0$. 

However, in ${\cal T}^{11}$, there is an additional term (the first term on the right hand side of equation~(\ref{eq:84.6c})), which actually is the $x$ component of momentum flux (momentum crossing the surface $x$ = constant per unit area per unit time). This is because the surface $x$ = constant is actually stationary in frame  ${\cal S}$, where the fluid is moving with a bulk velocity $v$ along x-axis. Thus through the surface $x$ = constant, per unit area per unit time, a transfer of  momentum, which is $v$ times the momentum density along direction $x$, or $v{\cal T}^{10}/c=\gamma^2 \left(\rho_0 +p\right){v^2/ c^2}$ takes place, which is the extra term seen in ${\cal T}^{11}$. 

The stress tensor ${\cal T}^{ij}$ in equations~(\ref{eq:84.6a}) and (\ref{eq:84.6a1}), in the case of radiation, can be understood in the same way. 
However ${\cal T}^{00}$ as well as ${\cal T}^{i0}$, in equations~(\ref{eq:84.4c}) and (\ref{eq:84.5c}) as well as in equations~(\ref{eq:84.4a}) and (\ref{eq:84.5a}), are not that easily explained.

The total energy contained in volume $V_0$ of the system, in rest frame ${\cal S}_0$ is 
\begin{eqnarray}
\label{eq:84.4d}
{\cal E}_0=\rho_0 V_0\:. 
\end{eqnarray}
The momentum of the system, 
in the rest-frame ${\cal S}_0$, is zero. 
\begin{eqnarray}
\label{eq:84.5d}
{P}_0=0\:. 
\end{eqnarray}

Seen from the lab frame ${\cal S}$, the ${\cal S}_0$ is moving with a velocity $v$ along the $x$-axis, therefore due to the Lorentz contraction, volume $V$ in  ${\cal S}$ is related to $V_{0}$ in rest frame ${\cal S}_0$ as
\begin{eqnarray}
\label{eq:84.18}
 V=\frac{V_0}{\gamma}
\end{eqnarray}
Then the total energy of the system in ${\cal S}$, from equation~(\ref{eq:84.4c}), is 
\begin{eqnarray}
\label{eq:84.4e}
{\cal E}={\cal T}^{00} V = \gamma V_0 \left(\rho_0 +p {v^2\over c^2}\right)\nonumber\\
= \gamma  \left({\cal E}_0 +p V_0{v^2\over c^2}\right)\:,
\end{eqnarray}
while the total momentum in ${\cal S}$, from Eq~(\ref{eq:84.5c}), is along x-axis with a magnitude 
\begin{eqnarray}
\label{eq:84.5e}
{P}= \frac{1}{c}{\cal T}^{10} V = \gamma\left({\cal E}_0 +pV_0\right) {v\over c^2}\:. 
\end{eqnarray}
In the case fluid comprises only radiation, then we get the energy momentum relations as 
\begin{eqnarray}
\label{eq:84.4f}
{\cal E}= \gamma {\cal E}_0 \left(1 +{v^2\over 3c^2}\right)\:,
\end{eqnarray}
\begin{eqnarray}
\label{eq:84.5f}
{P}= \frac{4}{3} {{\cal E}_0\over c^2} \gamma\: {v}\:. 
\end{eqnarray}

Here a question  arises that may be rather puzzling. Why does the expression for the energy ${\cal E}$ (equation~(\ref{eq:84.4e})) of the system in ${\cal S}$ contains pressure, $p$, besides ${\cal E}_0$? After all, the rest frame energy ${\cal E}_0$ of the system already contains not only the rest mass energy of the fluid constituents, but also their kinetic energy. For instance, in the ideal gas model of the fluid, the random motion of the gas molecules in ${\cal S}_0$ gives rise to the pressure in the fluid, but the kinetic energy due to this random motion is already included in $\rho_0$. Even in the case of radiation, $\rho_0$ comprises all the radiation energy density. Should not we then expect this closed system to have between ${\cal S}$ and ${\cal S}_0$ an energy relation ${\cal E}=\gamma {\cal E}_0$, instead of the that given by equation~(\ref{eq:84.4e}) or equation~(\ref{eq:84.4f})? 

Even more intriguing is the presence of pressure term in the expression (equation~(\ref{eq:84.5e})) for momentum of the system. After all, here we are not considering any rate of change of the momentum, where pressure might make an appearance, instead we are considering just the momentum of the system. One would have expected the momentum to be ${P}=\gamma{\cal E}_0 {v/ c^2}$, instead of that given by equation~(\ref{eq:84.5e}) or (\ref{eq:84.5f}), with ${\cal E}_0 {/ c^2}$ being the rest mass of the system.

In essence, the question posed here is: why do the energy and momentum of the electromagnetic field of a fluid system not behave as components of a 4-vector under a Lorentz 
transformation? As we will show below, these intriguing extra terms, especially in the expression for momentum of the system, historically speaking, are quite relevant, as these very terms are responsible for the famous, more than a century long, 4/3 problem in the electric mass of a classical, charged particle \cite{29,1,2}.
\section{How pressure contributes to the energy momentum of a system in motion -- a physical perspective} 
To understand the appearance of pressure in these two terms, and as well in equations~(\ref{eq:84.4c}) and (\ref{eq:84.5c}), the following two points need to be looked at carefully.
\begin{figure}[t]
\begin{center}
\includegraphics[width=\columnwidth]{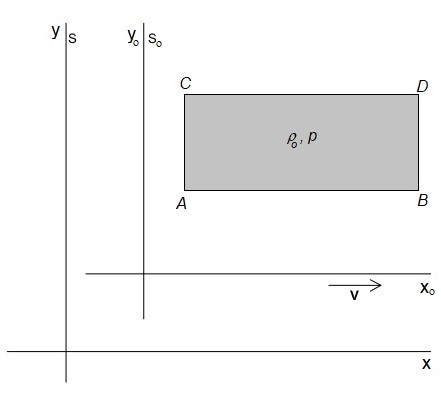}
\end{center}
\caption{A box $ABDC$ (the depth along z-axis suppressed in the diagram) containing perfect fluid, with mass density $\rho_0$ and pressure $p$ in its rest-frame $S_0$, moves with velocity ${\bf v}$ along the x-axis with respect to the lab frame ${\cal S}$.}
\end{figure}

Firstly, going from reference frame ${\cal S}_0$ to ${\cal S}$ there is a Lorentz contraction of the fluid by a factor $\gamma$, along the direction of motion. 
Now, Lorentz contraction of a system, when brought from one rest frame to another, arises because the acceleration applied simultaneously across the system in one frame (${\cal S}_0$) is not simultaneous for all points with respect to the other reference frame (${\cal S}$).  That gives rise to relative motion, as seen in ${\cal S}$, between different spacial points of the system and consequently a change in the dimensions of the system with respect to ${\cal S}$ occurs along the direction of acceleration \cite {83}. Lorentz contraction against pressure in a system adds to the energy content of the system.
This energy increment during Lorentz contraction is over and above 
the increase in the energy of the fluid due to the increase in its velocity and which would take place even in the absence of pressure.

Secondly there is an extra momentum in the fluid system, over and above that from the overall bodily movement $({\cal E}/c^{2})v$. In the reference frame ${\cal S}_0$, fluid constituents (molecules!) because of their random motion, continuously collide and get bounced off the container walls. During this collision process, there is no change in the energy of the reflected molecules in ${\cal S}_0$, as all collisions are assumed to be elastic for the perfect fluid. However, in frame ${\cal S}$, because the container walls  $AC$ and $BD$ (Fig.~1) are moving  along the $x$-axis (towards right, say), with speed $v$, the fluid molecules bounce off the wall $BD$ with reduced speed, and therefore with loss of kinetic energy, while those bouncing off the wall $AC$ gain kinetic energy. This way the molecules gain kinetic energy at the wall $AC$ and deliver that gain in energy at the wall $BD$. 

Let $A$ be the cross-section of container walls $AC$ and $BD$, in a direction normal to the $x$-axis. Then as the system moves along the $x$-axis with speed $v$, there is a continuous flow of energy {\em into} the fluid system at its left-sided surface of cross-section $A$ (due to the work being done against pressure $p$ at a rate ${ p\, A\, v}$), while at the same rate energy is flowing {\em out} of the fluid system at its right-sided surface. Even though there is no net change in the energy of the system, yet, due to this energy flows into and out of the system at two surfaces a distance $l/\gamma$ apart ($l$ being the corresponding distance in frame ${\cal S}_0$), there is a continuous transport of energy taking place from the left-sided surface to the right-sided surface within the fluid system, and this forms a part of the 
total momentum of the fluid system. 

We can calculate the energy $\cal E$ and momentum $P$ of the moving
fluid in the following way. Let us consider the change in 
energy of the system when it is taken from its state of rest to
that of a finite velocity along the $x$-axis, as seen in an inertial 
frame  ${\cal S}_0$. The implied transition of the system through successive 
`instantaneous' rest-frames would also mean, at each step, a change 
in Lorentz contraction of the system, as seen in  ${\cal S}_0$ \cite {83}.

The total rate of increase in energy of the system is given by
\begin{eqnarray}
\label{eq:84.10}
\frac{{\rm d}{\cal E}}{{\rm d}t}=\frac{{\rm d}{P}}{{\rm d}t}\,v+\frac{{\rm d}{\cal W}_{LC}}{{\rm d}t}\:.
 \end{eqnarray}
 Here the first term on the right hand side represents 
 the gain in energy due to bodily acceleration of the 
 system as a whole, and the second term is due to the work done against
the pressure during Lorentz contraction.

All dimensions along the x-axis shrink at
a rate ${\rm d}(\gamma ^{-1})/{\rm d}t$. Consider two opposite surfaces $AC$ and $BD$ of the box containing volume $V_0=A\,l$ of fluid, as seen in the rest-frame  ${\cal S}_0$ (Fig.~1). 
The rate of work being done on the volume $V_0$ due to Lorentz contraction against pressure $p$ is
\begin{eqnarray}
\label{eq:84.13}
\frac{{\rm d}{\cal W}_{LC}}{{\rm d}t}=-p V_0\,\frac{{\rm d}}{{\rm d}t}\left(
\frac{1}{\gamma}\right)\:.
\end{eqnarray}

Due to the movement of the system with a velocity $v$, the flow of energy per unit time {\em into} the fluid system at its left-sided surface is at a rate ${ p\, A\, v}$, while at the same rate energy is flowing {\em out} of the fluid system at its right-sided surface  
at a distance $l/\gamma$. This implies an energy flow at a rate 
$p \,A\, v\,l/\gamma$,
across the system, that amounts to a net momentum term, $p \,(V_0/\gamma)\,v\,/ c^{2}$, due to this energy flow.
 
The total momentum of the system of mass ${\cal E}/{c^{2}}$ moving with velocity ${v}$, therefore, is 
\begin{eqnarray}
\label{eq:84.15}
{P}=\frac{\cal E}{c^{2}}{v}+
\frac{p \,V_0}{\gamma}\frac{v}{c^{2}}\:.
\end{eqnarray}

Substituting equations~(\ref{eq:84.13}) and (\ref{eq:84.15}) in equation~(\ref{eq:84.10}), and after some simplifications we get,
\begin{eqnarray}
\label{eq:84.16}
\frac{{\rm d}}{{\rm d}t}\left({\cal E}+\frac{p V_{0}}{\gamma}\right)=
\left({\cal E}+\frac{p V_{0}}{\gamma}\right)\gamma ^{2}\frac{v}{c^{2}}
 \,\frac{{\rm d}v }{{\rm d}t}.
\end{eqnarray}
Integrating with time and noting that ${\cal E}_0$ is the 
energy in the rest frame ($v =0,\gamma=1$), we get
\begin{eqnarray}
\label{eq:84.17}
 {\cal E}=\gamma \,\left({\cal E}_0+p V_{0}\frac{v^{2}}{c^{2}}\right)\,.
\end{eqnarray}

Now substituting for $\cal E$ (equation~(\ref{eq:84.17})) in expression for momentum (equation~(\ref{eq:84.15})) we also get,
\begin{eqnarray}
\label{eq:84.20}
{P}=({\cal E}_0+p V_{0})\,\frac{\gamma v}{c^{2}}\,.
\end{eqnarray}
Equations~(\ref{eq:84.17}) and (\ref{eq:84.20}) are the expression (equations~(\ref{eq:84.4e}) and (\ref{eq:84.5e})), we aimed to derive for the
energy momentum of the perfect fluid system in a moving frame.

Dividing by $V$, we can write in terms of energy and momentum densities
\begin{eqnarray}   
\label{eq:84.19}
\frac{\cal E}{V}=\gamma^2 \,\left(\rho_{0}+p \frac{v^{2}}{c^{2}}\right)\,.
\end{eqnarray}
\begin{eqnarray}
\label{eq:84.21}
\frac{P}{V}=\gamma^2 (\rho_{0}+p) \frac{v}{c^{2}}\:, 
\end{eqnarray}
which are in agreement with equations~(\ref{eq:84.4c}) and (\ref{eq:84.5c}). Note the presence of pressure $p$ in equation~(\ref{eq:84.21}) even for non-relativistic bulk velocities.
\section{Energy momentum of the container forces}

Due to the pressure, even in the rest frame ${\cal S}_0$, the fluid would disperse unless  contained or confined in some way. There will then also be forces of the containing walls that exert pressure on the liquid in all such cases. We may not know the exact details of their nature or how these may get generated, and that may not even have any relevance for our purpose; all we can be sure about them is that these will keep 
the system in equilibrium by providing everywhere, a pressure equal and opposite to that of the liquid \cite{31}.

Therefore, their contributions to the energy and momentum of the system must also be 
equal and opposite to that calculated above for that due to the pressure in the liquid.  
Noting that any energy contribution of the  container forces is zero in the rest frame  ${\cal S}_0$, i.e. for $v =0$, we get
\begin{eqnarray}
\label{eq:84.17a}
\Delta{\cal E}=-\gamma\, p V_{0}\frac{v^{2}}{c^{2}}\:.
\end{eqnarray}
and 
\begin{eqnarray}
\label{eq:84.20a}
\Delta{P}=-\gamma \,p V_{0}\,\frac{ v}{c^{2}}\:.
\end{eqnarray}

Now, if we include these energy-momentum contributions to the total energy-momentum of the system, then we obtain
\begin{eqnarray}
\label{eq:84.17b}
 {\cal E}=\gamma \,{\cal E}_0\:,
\end{eqnarray}
\begin{eqnarray}
\label{eq:84.20b}
{P}={\cal E}_0\,\frac{\gamma v}{c^{2}}\:,
\end{eqnarray}
energy and momentum that transform like the components of a 4-vector.
\section{Electromagnetic energy-momentum tensor}
The electromagnetic energy-momentum tensor can be written as \cite{MTW73}
\begin{eqnarray}
\label{eq:84.223}
{\cal T}^{\mu\nu} = \frac{1}{4\pi} [ {\cal F}^{\mu\alpha}{\cal F}_{\alpha\beta}\:\eta^{\beta\nu} - \frac{1}{4} \eta^{\mu\nu}{\cal F}^{\alpha\beta}{\cal F}_{\alpha\beta}] \,,
\end{eqnarray}
where ${\cal F}^{\mu \nu }$ is the electromagnetic field tensor, given by \cite{MTW73}
\begin{eqnarray}
\label{eq:84.22}
{\displaystyle {\cal F}^{\mu \nu }
={\begin{bmatrix}0&E^{1}&E^{2}&E^{3}\\-E^{1}&0&B^{3}&-B^{2}\\-E^{2}&-B^{3}&0&B^{1}\\-E^{3}&B^{2}&-B^{1}&0\end{bmatrix}}.} 
\end{eqnarray}
We can use equations~(\ref{eq:84.223}), (\ref{eq:84.22}) along with (\ref{eq:84.3}), to express electromagnetic energy-momentum tensor in component form. 
Then we get
\begin{eqnarray}
\label{eq:84.24}
{\cal T}^{00}={\frac {1}{8\pi }}(E^{2}+B^{2})\,,
\end{eqnarray}
which is the electromagnetic energy density, 
\begin{eqnarray}
\label{eq:84.25}
{\cal T}^{0i} = {\cal T}^{i0} = {\frac {1}{4\pi}}(\mathbf {E} \times \mathbf {B})^i\,, 
\end{eqnarray}
which gives $c$ times the electromagnetic momentum density, and 
\begin{eqnarray}
\label{eq:84.26}
{\displaystyle {\cal T}^{ij}=\frac {-1}{4\pi }\left[E^{i}E^{j}+B^{i}B^{j}-{\frac {1}{2 }}\left(E^{2}+B^{2}\right)\delta^{ij}\right]}
\end{eqnarray}
is the Maxwell stress tensor. 

It should be noted that the stress tensor describes a negative pressure (tension) $(E^{2}+B^{2})/{8\pi }$ along the field lines and an equal but positive  pressure at right angles to the field lines \cite{MTW73,RI06}.
\subsection{Energy momentum of a charged capacitor in motion}
We consider a charged parallel plate capacitor, with a surface charge density $\sigma$, stationary in rest frame ${\cal S}_0$. The only non-zero component of the electric field is, $E_0=4\pi\sigma$ \cite{PU85}, along the plate separation, within the capacitor volume, but is zero outside. Magnetic field $B$ is zero throughout in frame ${\cal S}_0$. The electromagnetic field energy density, from equation~(\ref{eq:84.24}), is ${\cal T}^{0'0'}=E_0^2/8\pi=2\pi\sigma^2$ within the capacitor volume, but zero outside, while the momentum density is throughout zero,  in the rest frame ${\cal S}_0$.
\subsubsection{Motion normal to the plate surfaces}
We assume the plates to be normal to the x-axis and the only non-zero components of the electromagnetic field tensor in the rest frame ${\cal S}_0$ are ${\cal F}^{0'1'}=-{\cal F}^{1'0'}=E_0$, a prime on an index indicating it refers to the rest frame ${\cal S}_0$.
Seen from the frame ${\cal S}$, the charged capacitor is moving with a velocity $\bf v$, along the x-axis (Fig.~2). 

From a Lorentz transformation of the field tensor, ${\displaystyle {\cal F}^{\mu\nu}={\Lambda ^{\mu}}_{\alpha '}{\Lambda ^{\nu}}_{\beta '}{\cal F}^{\alpha ' \beta '},}$
it is easily seen that even in frame ${\cal S}$ the only non-zero field components again are ${\cal F}^{01}=-{\cal F}^{10}=E_0$, implying the electric field vector, which is along the $x$-axis, remains unchanged, $E=E_0$. Thus there is no change in the field energy density, $\rho=\rho_0$. But as the whole system is Lorentz contracted by a factor $\gamma$ along the x-axis, $V=V_0/\gamma$ and the total electromagnetic field energy, from equation~(\ref{eq:84.24}), as calculated in the frame ${\cal S}$ is 
\begin{eqnarray}
\label{eq:84.4g1}
{\cal E}= \frac{E_0^2}{8\pi}V=\frac{{\cal E}_0}{\gamma}=\frac{\rho_0 V_0}{\gamma} \:,
\end{eqnarray}
where ${\cal E}_0=\rho_0 V_0$ is the electromagnetic field energy in the rest frame ${\cal S}_0$. This reduction of energy by a factor $\gamma$ seems at odds \cite{RD88} with the standard expectation that the energy of a moving system should rather be higher by a factor $\gamma$. 
\begin{figure}[t]
\begin{center}
\includegraphics[width=\columnwidth]{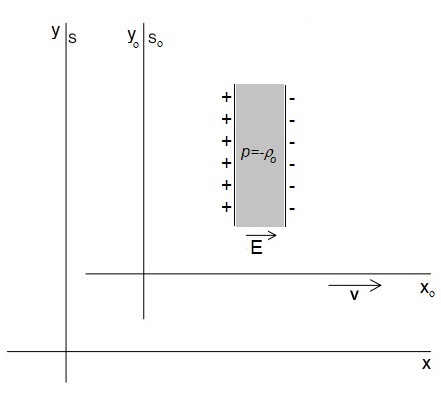}
\end{center}
\caption{A charged parallel plate capacitor, stationary in frame ${\cal S}_0$, moves with velocity ${\bf v}$, in a direction normal to the plate surfaces, in the lab frame ${\cal S}$. The energy density of the system in the rest frame ${\cal S}_0$ is $2\pi\sigma^2$. Also, there is an inward  force per unit area, $2\pi\sigma^2$ on each plate, implying an equivalent fluid with a negative pressure or tension, $p=-\rho_0$.}
\end{figure}

The magnetic field is zero everywhere, therefore, from equation~(\ref{eq:84.25}), the field momentum is zero, in frame ${\cal S}$ too.
\begin{eqnarray}
\label{eq:84.5g1}
{P}= 0. 
\end{eqnarray}
This could appear as a paradox with a nil electromagnetic momentum for a moving  system, having a finite electromagnetic energy stored inside, since a system with energy ${\cal E}$ and moving with a velocity 
${\bf v}$, is expected to possess a momentum ${\cal E}{\bf v}/c^2$. How come a moving system with a finite energy has, instead, a zero momentum?

In order to understand these puzzling aspects of the energy momentum transformations, we consider the capacitor system to be equivalent to a volume (between the capacitor plates) filled with a fluid of energy density  $\rho_0={\cal T}^{0'0'}=E_0^2/8\pi$ (equation~(\ref{eq:84.24})) in the rest frame ${\cal S}_0$. The momentum density of this system is zero (equation~(\ref{eq:84.25})) in ${\cal S}_0$.  The stress component ${\cal T}^{1'1'}=-E_0^2/8\pi$ (equation~(\ref{eq:84.26})) for the fluid, implies a  negative pressure (or tension), $p=-\rho_0$, along the x direction, which is consistent with an  {\em inward} force per unit area $2\pi\sigma^2$ on each plate \cite{PU85} (due to the electrical attraction between the plates of the capacitor). 

Then from equations~(\ref{eq:84.17}) and (\ref{eq:84.20}), we get  
\begin{eqnarray}
\label{eq:84.4g}
{\cal E}= \frac{{\cal E}_0}{\gamma} \:,
\end{eqnarray}
\begin{eqnarray}
\label{eq:84.5g}
{P}= 0. 
\end{eqnarray}

Thus we see that the expressions for energy and momentum in the electromagnetic fields of the capacitor (equations~(\ref{eq:84.4g1}) and (\ref{eq:84.5g1})), are identical to those derived for the fluid (equations~(\ref{eq:84.4g}) and (\ref{eq:84.5g})), which include contributions of the work done by the pressure (tension in this case) during a Lorentz contraction, and the momentum associated with a continuous energy flow taking place along the direction of motion, across the capacitor volume. 
 
\subsubsection{Motion parallel to the plate surfaces}
\begin{figure}[t]
\begin{center}
\includegraphics[width=\columnwidth]{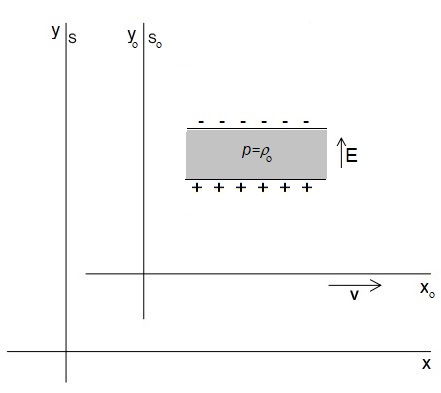}
\end{center}
\caption{A charged parallel plate capacitor, stationary in frame ${\cal S}_0$, moves  with velocity ${\bf v}$ along x-axis, a direction parallel to the plate surfaces, in the lab frame ${\cal S}$. The electric field is along the y-axis. In the lab frame ${\cal S}$, there is also a magnetic field (not shown here) along the $z$ direction, i.e., out of the plane of the paper.}
\end{figure}
We assume the plates oriented normal to the $y$-axis and the only non-zero components of the electromagnetic field tensor in the rest frame ${\cal S}_0$ are ${\cal F}^{0'2'}=-{\cal F}^{2'0'}=E_0$.
Then, in frame ${\cal S}$, with the capacitor moving along
the $x$-axis (Fig.~3), we find, using a Lorentz transformation, ${\displaystyle {\cal F}^{\mu\nu}={\Lambda ^{\mu}}_{\alpha '}{\Lambda ^{\nu}}_{\beta '}{\cal F}^{\alpha ' \beta '},}$ 
 that the only non-zero field components in frame ${\cal S}$ are ${\cal F}^{02}=-{\cal F}^{20}=\gamma E_0$ and ${\cal F}^{12}=-{\cal F}^{21}=\gamma E_0v/c$. This implies from equation~(\ref{eq:84.22}), an electric field $E=\gamma E_0$ along the $y$-axis and a magnetic field, ${B}=({\gamma v}/c)  E_0$, along the $z$-axis. 
With the volume reducing due to Lorentz contraction of plate dimensions along the $x$ direction by a factor $\gamma$ the total energy in the electromagnetic fields, from equation~(\ref{eq:84.24}), therefore, is 
\begin{eqnarray}
\label{eq:84.21a}
{\cal E}={\cal T}^{00} V=\gamma {\cal E}_{0} \left(1+\frac{v^2}{c^{2}}\right)\,.
\end{eqnarray}
From equation~(\ref{eq:84.25}), there is also a field momentum
\begin{eqnarray}
\label{eq:84.21b}
{P}=\frac{1}{c}{\cal T}^{10} V=2{\cal E}_{0}\gamma \,\frac{v}{c^{2}}\,,
\end{eqnarray}
along the $x$-axis.

The expression for energy seems at odds with the standard transformation formula, ${\cal E}=\gamma {\cal E}_{0}$, and so is the momentum formula, from the expected expression  
for the momentum, ${P}=({\cal E}/c^2){v}=\gamma ({\cal E}_{0}/c^2){v}$. 

Again, considering an equivalent fluid filling the volume between the capacitor plates, we note that in the rest frame ${\cal S}_0$, the energy density is $\rho_0=2\pi\sigma^2$, while the stress component ${\cal T}^{1'1'}=E_0^2/8\pi=2\pi\sigma^2$ (equation~(\ref{eq:84.26})), implying along the $x$ direction, a pressure $p=\rho_0$. Unlike in the case of a negative pressure or tension along the field lines discussed in the previous section, and which represented the force of attraction between oppositely charged capacitor plates, the genesis of a positive pressure, normal to the field lines, might not be so evident here. 
Actually, there are electromagnetic forces of repulsion, on similar charges {\em within} each plate, 
along the plate surface \cite{15}. A positive pressure is a representation of these forces of repulsion and contributes accordingly to the energy-momentum of the system moving along the $x$-axis.

Then from equations~(\ref{eq:84.17}) and (\ref{eq:84.20}), we get  
\begin{eqnarray}
\label{eq:84.21c}
{\cal E}=\gamma {\cal E}_{0} \left(1+\frac{v^2}{c^{2}}\right)\,,
\end{eqnarray}
and
\begin{eqnarray}
\label{eq:84.21d}
{P}=2{\cal E}_{0}\gamma \,\frac{v}{c^{2}}\,,
\end{eqnarray}
which are identical to the expressions obtained for the energy and 
momentum equations~(\ref{eq:84.21a}) and (\ref{eq:84.21b}) of the electromagnetic fields in frame ${\cal S}$. Equations~(\ref{eq:84.21c}) and (\ref{eq:84.21d})) include contributions of the work done by the pressure during a Lorentz contraction, and the momentum associated with a continuous energy flow taking place along the direction of motion, across the capacitor volume. Of course, one obtains exactly the same expressions for the energy momentum of the system in an alternate way to calculate the work done, by directly exploiting the forces of repulsion within the capacitor plates, though their evaluation could relatively be quite involved, computationally \cite{15}.
\subsection{Energy momentum of a charged particle in motion}
\begin{figure}[t]
\begin{center}
\includegraphics[width=\columnwidth]{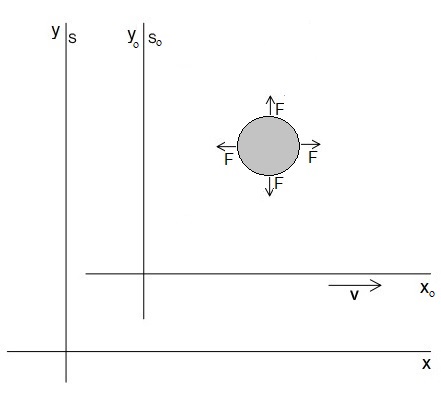}
\end{center}
\caption{Energy momentum contribution of the self-repulsion force of a charged spherical shell can be analyzed, by treating it equivalent to a spherical container filled with a ``perfect fluid'' comprising electromagnetic radiation (virtual photons!), exerting a uniform pressure in all directions, in the rest frame ${\cal S}_0$. The sphere containing liquid, moves with velocity ${\bf v}$ along the x-axis with respect to the lab frame ${\cal S}$.}
\end{figure}
We consider a charge $e$, distributed uniformly in a thin spherical shell, $\Sigma$, of radius $\epsilon$, with a surface charge density, $\sigma=e/4\pi \epsilon^{2}$. Due to the mutual force of electrostatic repulsion between various constituents of the charge distribution, 
there is a radially outward surface force, $2 \pi \sigma ^{2}$, per unit area \cite{PU85}, 
in the rest-frame ${\cal S}_0$. 
The total energy, ${\cal E}_{0}$, of the system is the self-potential energy of the charged sphere, evaluated from a double surface integral over $\Sigma$, as 
\begin{eqnarray}
\label{eq:84.23}
{\cal E}_{0}= \frac{1}{2} \int\!\!\int \frac{\sigma
({\bf x})\sigma ({\bf x}')}{|{\bf x}-{\bf x}'|}\,{\rm d}\Sigma\,{\rm d}\Sigma' ={e^{2}}/{2\epsilon}.
\end{eqnarray} 
With the energy ${\cal E}_{0}$, we can associate a mass, ${\cal E}_{0}/c^{2}$, 
called the electromagnetic mass of the system, in rest frame ${\cal S}_0$. Since the charge is stationary in ${\cal S}_0$, the momentum is zero. Seen from another frame ${\cal S}$, the charge is moving with velocity $v$ along the x-axis. Our aim here is to compute the energy and momentum of the system in frame ${\cal S}$, comparative to the values in the rest-frame ${\cal S}_0$. 

The radially outward repulsive force, $2 \pi \sigma ^{2}$ per unit area, distributed uniformly on the charged spherical surface can be considered mathematically equivalent to a sphere (Fig.~4), filled with  electromagnetic radiation (virtual photons!), 
with a uniform outward pressure 
\begin{eqnarray}
\label{eq:84.23a}
p=2 \pi \sigma ^{2}\,,
\end{eqnarray}
and, from equation~(\ref{eq:84.23}), an energy density
\begin{eqnarray}
\label{eq:84.23b}
\rho_0=\frac{{\cal E}_{0}}{V_0}=\frac{{e^2}}{2\epsilon V_0}= 6\pi \sigma ^{2}\,.
\end{eqnarray}
The system thus obeys the relation $p= \rho_0/3$, as in equation~(\ref{eq:84.4.1}), justifying thereby equivalence of the charged particle, here, to a sphere filled with electromagnetic radiation with energy density $\rho_0$ and pressure $p$.

Then, in order to find out the energy momentum of the charged distribution, we can use directly the formulas derived earlier, (equations~(\ref{eq:84.4f}) and (\ref{eq:84.5f})), for the fluid comprising radiation, to write  
\begin{eqnarray}
\label{eq:84.4h}
{\cal E}= \gamma {\cal E}_0 \left(1 +{v^2\over 3c^2}\right)\:,
\end{eqnarray}
\begin{eqnarray}
\label{eq:84.5h}
{P}= \frac{4}{3} {{\cal E}_0\over c^2} \gamma\: {v}\:. 
\end{eqnarray}
This gives the energy and momentum of a charged particle, moving with a uniform velocity $v$, which agrees with the energy and momentum in the fields of the system derived in Appendix. Of course, one can also arrive at the same expressions, alternatively, by direct evaluations of the $x$ components of the forces of self-repulsion on each infinitesimal element of the charged spherical shell \cite{15}.
\section{Discussion}
We demonstrated above how the pressure term in a perfect fluid  
makes additional contribution to the energy-momentum of the system. During the Lorentz contraction of the system, when the system is accelerated from one inertial frame to another, work is done against pressure, which makes a finite contribution to the total energy of the system. Also, when a fluid system has a bulk motion, work done against pressure is positive on one side of the system, while on the opposite side, an equal amount work is being done {\em by} the system and this implies a continuous transport of energy from one side of the system to the other, thereby making a finite contribution to the momentum of the system.  These contributions of pressure to the energy-momentum of the system, already present in the relativistic formulation of a perfect fluid, have remained shrouded in mystery, from a physical perspective. 
Of course, if the liquid is contained and one includes the contribution of the container pressure too, then these pressure terms get cancelled in the expressions for energy-momentum, which then behaves as a 4-vector under a Lorentz transformation. However, in systems, where the liquid does not have an external container, and may be self-contained, like in the relativistic description of a spherical star or the cosmological model of the Universe expansion \cite{MTW73,SC85}, one has to use the fluid equations, including the pressure contribution to the energy-momentum of the system.

In the case of electromagnetic systems, where electromagnetic forces might be prevailing, the standard formulas for energy-momentum of the electromagnetic fields already contain the  contributions of the electromagnetic forces to the energy-momentum of the system. 
However, there has been a persistent opinion that the energy and momentum in the electromagnetic field of a moving charge should behave as components of a 4-vector under a Lorentz transformation. With this in mind, modifications in definitions of energy and momentum have been proposed,  in the case of ``bound fields'' \cite{Rohrlich60};  these modified definitions have  made their appearance even in the standard textbooks \cite{1,2}. However, as was recently  demonstrated \cite{31}, there is no need for separate definitions for energy-momentum of bound fields (fields associated with charges) and free fields (electromagnetic waves no longer tied to the charges, responsible for their generation). All we need to remember in the case of systems involving charges is that there might be contributions of 
the stabilizing forces (Poincar\'{e} stresses \cite{34}) too, to the energy and momentum of the total system, which would neutralize those ``unwanted'' terms. It has been demonstrated that the  stabilizing forces, responsible for retaining the charges on individual capacitor plates in spite of the forces of repulsion between charges along the plate surfaces and also for keeping the  plates apart against the perpendicular force of attraction between them, do explain successfully \cite{58} the null results of the famous Trouton-Noble experiment \cite{59}.

Therefore, for ``eliminating unwanted terms'', like, for instance, the intriguing factor of 4/3 in the electromagnetic mass of a charge, there is absolutely no reason for adopting a change in the conventional formulas. Further, both approaches are not equivalent, as sometimes suggested in the literature \cite {Teukolsky96}. 
At least in the case of relativistic fluid models, no such drastic modifications in the definitions of the energy-momentum of a system have ever been proposed, and the physics remains the same in the electromagnetic case too.  
\section{Conclusions}
Starting from the simple expression for energy-momentum density of a perfect fluid systems in its rest frame, we first derived the expression for energy-momentum density in another inertial frame in which the fluid moves, with a constant velocity. We thereby showed the presence of a couple of pressure-dependent ``mysterious'' terms, one arising because of the work done by pressure against the Lorentz contraction of the system, and the second adding to the momentum density from a rate of transport of energy from one part of the system to another due to pressure, all {\em within} the system itself. We resolved the presence of the famous, more than a century old, intriguing factor of 4/3, in the electrical mass within the classical model of an electron, or any other electric charge, assumed to be of a spherical shape in its rest frame. 

Further, we showed that in a electrically charged parallel-plate capacitor moving normal to plate surfaces, where not only is the electromagnetic momentum nil, even the field energy decreases with an increase in the system velocity, it is the contribution of these very terms to the energy-momentum of the electromagnetic system that explains their apparently paradoxical behaviour. 
It was emphasized that there is no need for a modified definition of energy-momentum for bound fields or electromagnetic systems involving forces due to electric charges, contrary to some suggestions in the literature, in the same way as there has been no need for any similar modifications in the case of relativistic perfect fluids with pressure, and the physics remains the same in the electromagnetic case too.  
\section*{Appendix}
\section*{Energy momentum in the electromagnetic fields of a charged particle in motion}
We assume the charged particle to be a thin spherical shell of radius $\epsilon$ of a uniform surface charge density, with total charge $e$, stationary in a frame ${\cal S}_0$, called the rest frame (Fig.~5). The electric field is zero inside the shell, and is radial outside the shell, with magnitude $E_0=e/r^2$. 
\begin{figure}[t]
\begin{center}
\includegraphics[width=\columnwidth]{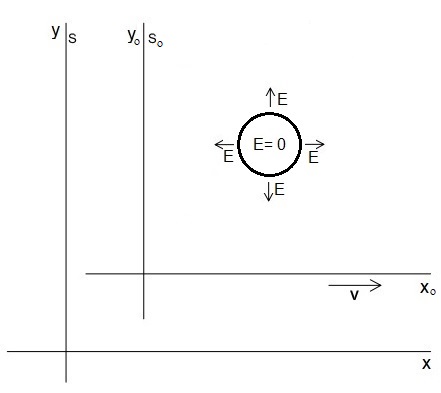}
\end{center}
\caption{A uniformly charged, thin spherical shell, has only a radial electric field outside the shell, but no fields inside the shell, as seen in the rest frame ${\cal S}_0$. The charged shell moves with velocity ${\bf v}$ along the x-axis with respect to the lab frame ${\cal S}$, where both electric and magnetic fields exist outside the spheroid-shaped shell, but nil fields on the inside.}
\end{figure}
  
The volume integral of the energy density in rest frame ${\cal S}_0$ is easily calculated to be 
\begin{eqnarray}
\label{eq:84.29}
{\cal E}_{0}=\int_\epsilon^\infty \frac{E_0^{2}}{8\pi }\,{\rm d}V_0 = \frac{e^2}{2\epsilon}\:.
\end{eqnarray}
The momentum is zero as the magnetic field is zero throughout

The volume integrals of the energy and momentum densities of the electromagnetic fields of the  charge in frame ${\cal S}$, where it is moving with a velocity $\bf v$, say along x-axis, are then computed, making use of equations~(\ref{eq:84.24}) and (\ref{eq:84.25}), as 
\begin{eqnarray}
\label{eq:84.30}
{\cal E}=\int \frac{E^{2}+B^{2}}{8\pi }\,{\rm d}V\:,
\end{eqnarray}
\begin{eqnarray}
\label{eq:84.31}
{\bf P}=\int \frac{{\bf E} \times {\bf B}}{4\pi c}\,{\rm d}V\:,
\end{eqnarray}
where $\bf E$ and $\bf B$ are the electric and magnetic fields of the
uniformly moving charge.  

These integrals can be evaluated in a simple way by noting that ${\bf B}=\mbox{\boldmath $\bf v$} \times {\bf E}/c$, and using the transformation relations  between ${\cal S}$ and ${\cal S}_0$, for the electromagnetic fields as well as for the volume element as 
\begin{eqnarray}
\label{eq:84.32}
E_{\|}&=&E_{0\|}\,,\\ 
E_{\bot}&=&\gamma E_{0\bot}\,,\\ 
{\rm d}V&=&{\rm d}V_{0}/\gamma\:. 
\end{eqnarray}

From that we get 
\begin{eqnarray}
\label{eq:84.33}
{\cal E}=\int \frac{E^{2}_{0\|}+(\gamma^{2}+\gamma^{2}v^2/c^{2}) E^{2}_{0\bot}}{8\pi\gamma}\,{\rm d}V_0\:.
\end{eqnarray}

Now because of the circular-cylindrical symmetry of the system about the x-axis, the direction of motion, we have
\begin{eqnarray}
\label{eq:84.34}
 \int E^{2}_{0\bot}dV_0=\frac{2}{3}\int E_0^{2}\,{\rm d}V_0\:.
\end{eqnarray}
Then, after integration, we get for the field energy in ${\cal S}$
\begin{eqnarray}
\label{eq:84.35}
{\cal E}=\gamma {\cal E}_{0}\left(1+\frac{v^{2}}{3c^2}\right)\:.
\end{eqnarray}
In the same way, we have an expression for the field momentum 
\begin{eqnarray}
\label{eq:84.36}
{\bf P}=\int \frac{\gamma^{2}{\bf v} E^{2}_{0\bot}}{4\pi c^2 \gamma}\,{\rm d}V_0\:,
\end{eqnarray}
which, after integration, yields
\begin{eqnarray}
\label{eq:84.37}
{\bf P}=\frac{4}{3}\,{\cal E}_{0} \frac{\gamma \bf v}{c^{2}}\:.
\end{eqnarray}
\section{Declarations}
The author has no conflicts of interest/competing interests to declare that are relevant to the content of this article. No funds, grants, or other support of any kind was received from anywhere for this research.
{}
\end{document}